\documentclass[10pt,twocolumn]{wlscirep}
\pdfoutput=1 
\usepackage{amssymb,amsmath}
\usepackage{graphicx}
\usepackage{xcolor}
\usepackage{subfiles}
\usepackage{subfigure}
\usepackage{hyperref}
\usepackage{listings}
\usepackage{siunitx}
\usepackage{gensymb}
\usepackage{textcomp}
\usepackage{xspace}
\usepackage{cleveref}
\usepackage{multirow}
\usepackage{float}
\usepackage[normalem]{ulem}
\usepackage{caption}
\usepackage{lineno}
\usepackage{etoolbox}

\definecolor{natureblue}{rgb}{0.282, 0.451, 0.643}
\definecolor{naturebluedark}{rgb}{0.236, 0.329, 0.533}

\hypersetup{
    unicode=false,        
    pdftoolbar=true,        
    pdfmenubar=true,        
    pdffitwindow=false,     
    pdfstartview={FitH},    
    pdfauthor={none},     
    colorlinks=true,       
    linkcolor=naturebluedark,          
    citecolor=naturebluedark,        
}

\newcommand{\ket}[1] {| #1 \rangle}

\title{Comparing Shor and Steane Error Correction Using the Bacon-Shor Code}
\author[1,2,*]{Shilin Huang}
\author[1,2,3,4]{Kenneth R. Brown}
\author[1,3,$\dagger$]{Marko Cetina}

\affil[1]{Duke Quantum Center, Duke University, Durham, NC 27701}
\affil[3]{Department of Physics, Duke University, Durham, NC 27708}
\affil[2]{Department of Electrical and Computer Engineering, Duke University, Durham, NC 27708}
\affil[4]{Department of Chemistry, Duke University, Durham, NC 27708}

\affil[*]{Present Address: Department of Applied Physics, Yale University, New Haven, CT 06511}
\affil[$\dagger$]{Email: marko.cetina@duke.edu}

\begin{abstract}
\textbf{Quantum states can quickly decohere through interaction with the environment.  Quantum error correction is a method for preserving coherence through active feedback. Quantum error correction encodes the quantum information into a logical state with a high-degree of symmetry. Perturbations are first detected by measuring the symmetries of the quantum state and then corrected by applying a set of gates based on the measurements.  In order to measure the symmetries without perturbing the data, ancillary quantum states are required. Shor error correction uses a separate quantum state for the measurement of each symmetry. Steane error correction maps the perturbations onto a logical ancilla qubit, which is then measured to check several symmetries simultaneously. Here we experimentally compare Shor and Steane correction of bit flip errors using the Bacon-Shor code implemented in a chain of 23 trapped atomic ions.  We find that the Steane error correction provides better logical error rates after a single-round of error correction and less disturbance to the data qubits without error correction.  
} 
\end{abstract}

\begin{document}
\flushbottom
\maketitle

\section{Introduction}
Quantum computers promise to solve computationally difficult cryptographic and chemistry problems using billions or trillions of quantum gates \cite{GidneyQuantum2021, microsoft2022}. Current quantum devices are limited to hundreds of quantum gates and many believe the path forward is quantum error correction.  Quantum error correction and the theory of fault tolerant quantum computation enable us to make reliable qubits and gates from unreliable components.

In the last few years, multiple experiments have demonstrated key elements of quantum  error correction. Highlights include improved state preparation and measurement error \cite{EganNature2021,NguyenPPAppl2021}, extended coherence times \cite{SivakNature2023}, transversal logical gates \cite{EganNature2021,RyanAndersonPRX2021,RyanAndersonArxiv2022,PostlerNature2022}, repeated rounds of syndrome extraction \cite{RyanAndersonPRX2021,KrinnerNature2022,SundaresanNatComm2023,AcharyaNature2023,ZhaoPRL2022}, magic-state injection \cite{PostlerNature2022}, improved performance with code distance \cite{AcharyaNature2023}, and encoded circuits with multiple logical qubits \cite{bluvstein2023logical}. No experiment has demonstrated an improvement of all aspects for an encoded qubit relative to a physical qubit. These early experiments allow us to test our assumptions not only about quantum error correction protocols, but also the underlying noise models.  These tests are useful scientific studies that help us chart the path towards useful and scalable quantum error correction.

Quantum error correction encodes logical qubits into multiple physical qubits.  The encoded space has a number of symmetries and detectable errors change the value of the symmetry. Stabilizer codes have binary symmetries corresponding to the $+1$ and $-1$ eigenvalues of the stabilizer operators, which are a commuting group of $n$-qubit Pauli operators \cite{gottesman1997stabilizer}. The codespace is typically defined as the $+1$ symmetry space. We do not measure the eigenvalues of all stabilizers, but only a subset that generates the whole space.  The error syndrome is a bit string where the 1's indicate that the state is now in the $-1$ eigenspace of that stabilizer generator.

\section{Shor versus Steane}


\begin{figure*}
   \includegraphics[width=\linewidth]{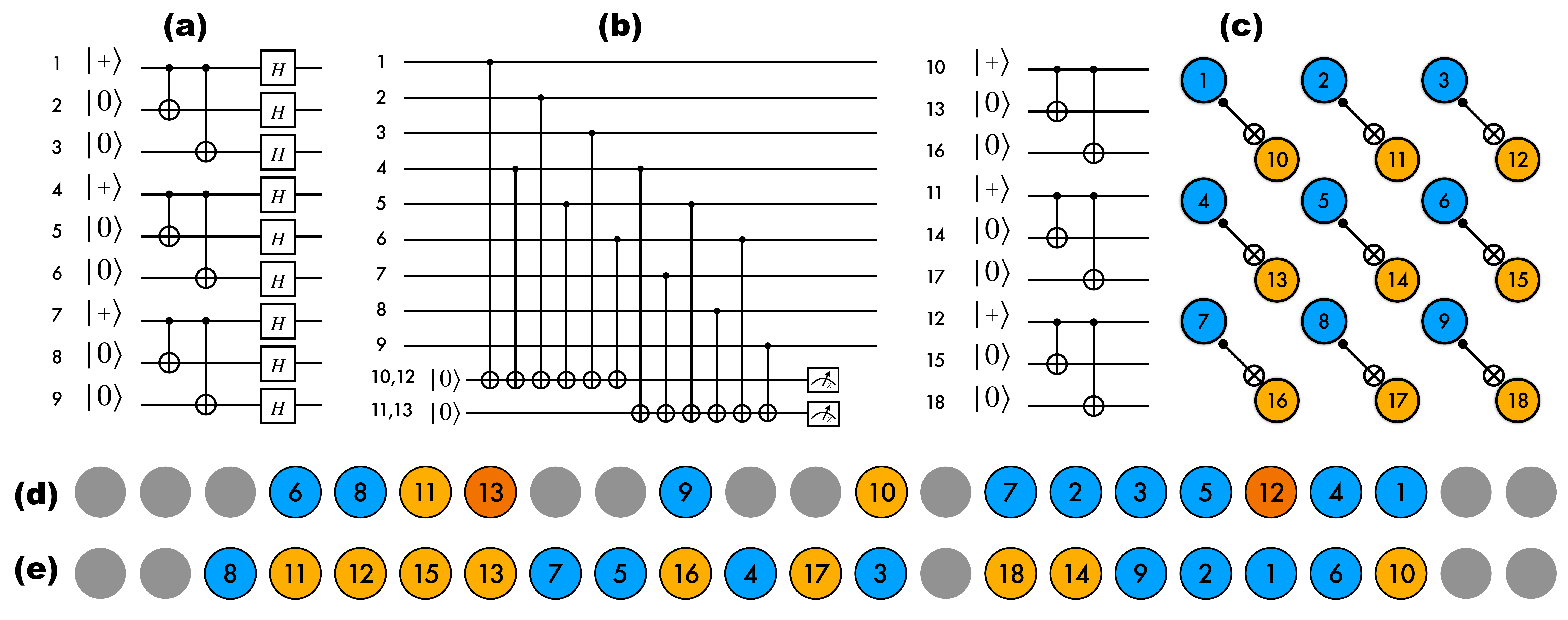} 
   \caption{(a) Bacon-Shor code can be encoded into the $\ket{0_L}$ state by preparing three copies of a three-qubit GHZ entangled state.  (b) Shor-style syndrome checks can be done with single-qubit ancilla. To repeat the measurement, we use a separate set of ancilla qubits.  (c) Steane-style syndrome extraction involves preparing a $\ket{+_L}$ state (circuit on left) and then performing a transversal CNOT between the data qubit and the ancilla qubit. (d) Ion positions in the 23 qubit ion used for Shor-style syndrome extraction: the logical qubit is encoded in ions labeled 1-9 and the ions labeled 10 and 11 (12 and 13) are two ancilla qubits for $S_1$ ($S_2$) (e) Ion positions in the 23 ion chain used for Steane-style syndrome extraction: the logical data qubit is encoded in ions labeled 1-9 and the logical ancilla qubit is encoded in ions labeled 10-18.}
   \label{fig:circuitsandions}
\end{figure*}

A critical part of a quantum error correction procedure is measuring the error syndrome. We refer to Shor methods as methods that measure the error syndrome stabilizer generator by stabilizer generator. Traditionally this requires the preparation of an entangled cat state as an encoded ancilla qubit for each generator \cite{ShorCatState}. For some codes a single qubit is an effective ancilla qubit \cite{FowlerSurfaceCodeThresh2009,TomitaLowDSC2014, li2018direct} and for any code a flag qubit system can be used instead of entangled ancilla qubits \cite{ChaoPRXQuantum2020}. Knill and Steane present alternative methods where logical qubits in the same code are used to determine the error syndrome in one or two steps \cite{SteaneSteaneEC1997,KnillKnillEC2005}.  Many other methods are possible \cite{HuangPRL2021} especially when considering the interplay between noisy syndrome bits and repeated measurements \cite{LaiPRA2017}.

Here we study the [[9,1,3]] Bacon-Shor code.  The Bacon-Shor code is a subsystem code and the canonical method to determine its error syndrome is to measure weight-2 checks which are then decoded to determine the syndromes \cite{BaconBaconShor2006,aliferis2007subsystem, PoulinPRL2005}. In a subsystem code some symmetries are allowed to freely fluctuate and are referred to as gauges. The weight-2 checks are the gauges of the code that can be used to determine the four-stabilizer generators. An alternative approach is to measure the four stabilizer generators directly with single qubit ancilla \cite{LiBareAnc2017}.  This method depends on a specific ordering of the circuit which follows the underlying gauges \cite{LiPRX2019}. From a broad perspective both of these methods fall under the category of Shor quantum error correction. Measuring the full-syndrome requires 24-48 two qubit gates between the data qubits and syndrome qubits in an adaptive approach \cite{li2018direct}.

A Steane or Knill approach requires only 18 two qubit gates for the Bacon-Shor code to touch the data or teleport the data, respectively. For any code, when gate errors dominate it is expected that Steane or Knill will outperform Shor codes. The cost is the preparation of the complicated ancilla qubits.  For the case of the distance-3 Bacon-Shor code, the logical state can be prepared fault-tolerantly using unitary preparation \cite{EganNature2021}. This is a key advantage for small circuits and greatly benefits a Steane style error correction circuit.

\section{Experiment Methods}

Our experiment tests the theoretical assumption that Steane-style syndrome extraction will cause less disturbance to the logical qubits than a Shor-style syndrome extraction using an ion trap quantum computer. We perform this test by examining a single-round of syndrome extraction followed by determining the logical state of the data with an error corrected measurement. We use a single chain of ions and a single round of measurement and limit our interrogation to measuring the $X$ error syndrome ($Z$ stabilizer generators). A full round of syndrome extraction would require measuring both the $X$ and the $Z$ syndrome.

The experimental system, based on Ref. \cite{EganNature2021}, consists of a near-equispaced linear chain of $^{171}$Yb$^+$ ions storing ground-state hyperfine qubits. The ions are held in a Sandia HOA-2.1.1 trap housed in a room-temperature UHV chamber and are manipulated using a 355-nm Raman process driven by single global laser beam together with up to 32 individually-addressed beams. The state of each qubit is read out using an array of photomultiplier tubes coupled to multimode fibers.

To allow a direct comparison between fault-tolerant Shor and Steane syndrome extractions, we extended our experimental system from 15 to 23 ions. The increase in the number of ions leads to a sharp increase in their axial motion and the attendent errors in individual addressing \cite{CetinaPRXQuantum2022}. To counter these errors, we confined the ions to decrease the chain spacing from 4.4 $\mu$m to 3.8 $\mu$m and tailored our gates to suppress  coupling to the chain's axial motion, as described in the Supplementary Material.

The circuits for the experiments and the qubit to ion mapping are shown in Fig. \ref{fig:circuitsandions}. The CNOTs and Hadamard gates are decomposed in a standard way into native M\o lmer-S\o rensen gates and single qubit $X$ and $Y$ rotations \cite{EganNature2021}. Details of the physical circuit model can be found in the Supplementary Material.

\section{Experiment Results}

Our baseline experiment is a Shor-style syndrome extraction of the two Z stabilizers following the  previous work \cite{EganNature2021}. The error-corrected SPAM error for $\ket{0_L}$ of $0.20^{+0.27}_{-0.14}\%$ is comparable with the previous result of $(0.23\pm 0.13)\%$, demonstrating our ability to control addressing errors in our larger system.

The Shor-style error correction procedure for $X$ errors involves measuring two weight-6 $Z$ stabilizer generators, $S_1$ and $S_2$. A fault-tolerant procedure requires measuring the checks multiple times due to logical errors.  Our adaptive protocol \cite{LiBareAnc2017} is to measure the stabilizer generators once if the error-free syndrome is detected and to measure twice in the other case. This protocol will be fault-tolerant for any distance 3 quantum error correction code. We did not perform mid-circuit measurements in this experiment, so we simulate this procedure by running circuits with one or two-sets of stabilizer measurements.

\begin{figure*}
   \centering
   \includegraphics[width=\textwidth]{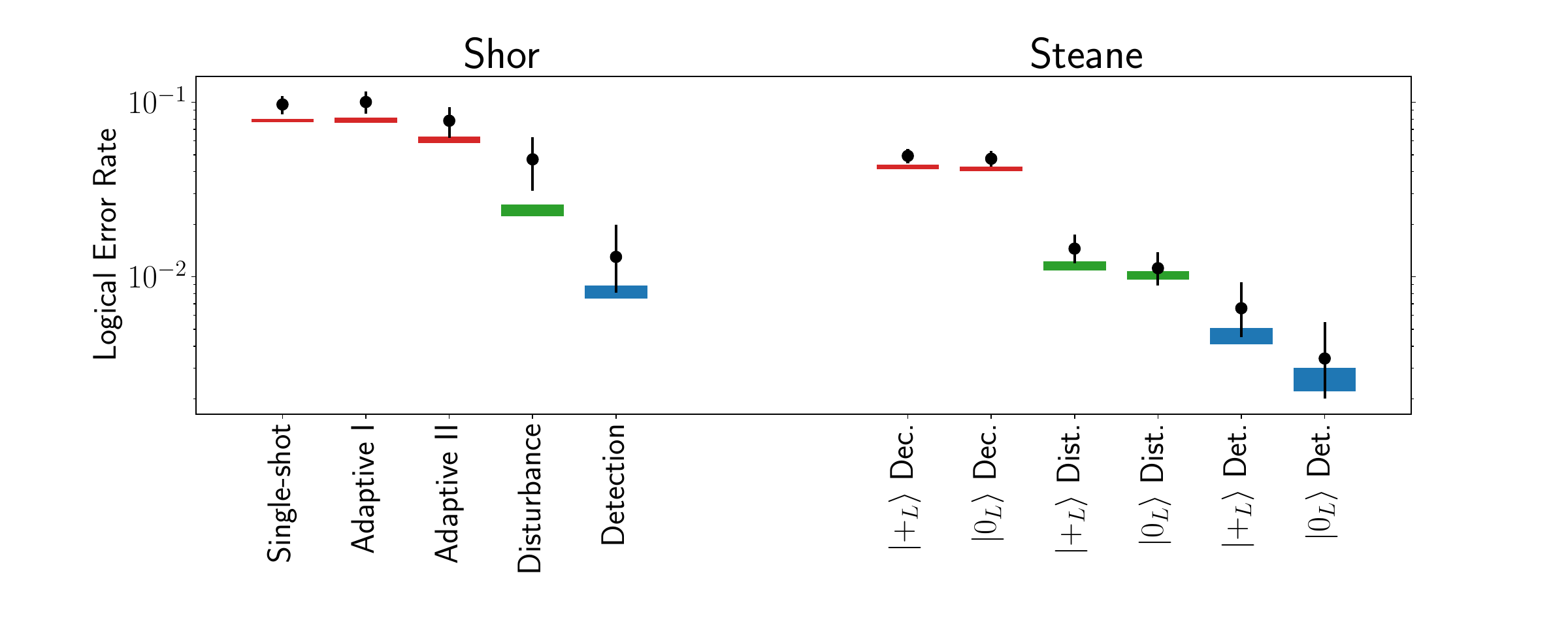}
    \caption{    
    Comparison of Shor-style and Steane-style error-correction protocols with the ancilla in the $|0_L\rangle$ and $|+_L\rangle$ logical states. Decoder (Dec., red) corrects the logical qubit with syndrome information before the final correction based on logical qubit  measurement. Disturbance (Dist., green) only corrects based on the logical qubit measurement. Detection (Det., blue) only keeps the data when no errors are detected on the ancilla qubits. 
    The Shor disturbance measurement reflects only a single round of syndrome measurements. The black dots are experimental data and the bars are numerical simulations. The errors correspond to 95\% confidence intervals. The simulations are based on error characterization experiments and are not adjusted for the logical qubit experiments.
    } 
   \label{Fig:ResultsBarChart}
\end{figure*}

We find a logical error rate based on correcting off the measured syndrome and then performing error correction on the measurement of $(10.0\pm1.5)\%$ (second column in Fig. \ref{Fig:ResultsBarChart}). We test a non-fault tolerant single shot procedure where we correct on the first round of stabilizers and find a similar overall logical error rate of $(9.7\pm1.2)\%$ (first column in Fig. \ref{Fig:ResultsBarChart}). A close inspection of the circuit reveals that the single-shot error correction fails when a data error occurs between the measurement of the two-stabilizers. Understanding these {\em internal errors} is critical for reducing the measurement overhead when the gates an measurements are faulty \cite{DelfosseIEEETransIT2022} In our improved adaptive decoder, we accept the syndromes $00$,$10$, and $11$ on the first round. If these syndromes have internal errors, then at least two single-qubit errors have occurred and we have already used up our fault budget; careful considerations of fault budgets can be used to reduce measurement overhead for many codes. \cite{TansuwannontQuantum2023,chao2018quantum,chamberland2018flag,delfosse2020short}. The improved adaptive decoder achieves a logical error rate of $(7.8\pm1.6)\%$ (third column in Fig. \ref{Fig:ResultsBarChart}).

Our Steane-style error correction experiment involves the prepartion of the data qubit in the $\ket{0_L}$ state and the syndrome qubits in the $\ket{+_L}$ state. A transversal CNOT between the data qubits and the syndrome qubits allows X errors in the preparation of the data qubits to be mapped to the syndrome qubits (\ref{fig:circuitsandions}c). One way to measure the negative impact of the syndrome circuit on the code is to measure the logical state and correct without any syndrome information. We find that the disturbance of the data qubits is significantly less for Steane errorr-correction than for the Shor case with both one and two-rounds of syndrome extraction.

For arbitrary data states, the logical ancilla state should be prepared in the $\ket{+_L}$ state, which should yield no back-action on the data qubits. For the $\ket{0_L}$ state, we can also use another $\ket{0_L}$ state as ancilla.  We test both ancilla states and find a logical error on the data qubit of $(4.93\pm0.49)\%$ for a $\ket{+_L}$ ancilla (column 6 in Fig. \ref{Fig:ResultsBarChart}) and a $(4.75\pm0.5)\%$ a $\ket{0_L}$ (column 7 in Fig. \ref{Fig:ResultsBarChart}) after we correct based on the ancilla qubit measurement.

We contrast our work to the recent neutral atom experiment~\cite{bluvstein2023logical}. In this experiment, $5$  logical qubits in a distance 3 color-code are encoded non-fault-tolerantly using a unitary circuit.  These logical qubits are then checked for $X$ errors using a Steane-style circuit onto another set of 5 logical qubits that are also not fault-tolerantly prepared.  This second set acts as flags which allows  for the postselection of errors. After post-selection, in a stochastic Pauli depolarizing error model with probability $p$, the data qubit will be in the wrong logical state with probability $O(p^2)$.

In our work, unitary preparation is fault-tolerant and two-errors are required for the logical ouput to fail. When we use a Steane-style circuit with post-selection, our state will be in the wrong logical state with probability $O(p^3)$ for a probabilistic error model.  Our state post-selected on errors measured when the ancilla qubit is prepared in the $\ket{+_L}$ state is  $0.66_{-0.21}^{+0.27}\%$ (column 10 in Fig. \ref{Fig:ResultsBarChart}) which outperforms the Shor-based post-selection (column 5 in Fig. \ref{Fig:ResultsBarChart}). If we use an ancilla in the $\ket{0_L}$ state, we can also post-select on logical errors.  This further improves the logical fidelity output to $0.34_{-0.14}^{+0.21}\%$ (last column in Fig. \ref{Fig:ResultsBarChart}), which is an improvement over not checking the error syndrome at the cost of a 34\% rejection rate. 

\section{Simulation and Perspective}
We developed a simulation of the experiment based on unitary evolution of the full wavefunction where errors are added probabilistically as unwanted unitary transformations. The error model is derived from experimentally measured errors in state preparation and measurement, single qubit gate errors, two-qubit gate errors including crosstalk, and the effect of axial heating of the ion chain on effective Rabi frequencies. The full details are in the Supplementary Material. 

The simulation is not fit to the logical error data and the parameters are determined from independent characterization and calibration experiments. This results in non-identical errors on different qubits in the chain. We see broad agreement between the simulated and the experimental data across the measured parameters.

The simulation allows us to determine the best steps forward for reducing logical error. We find that the dominant source of error is the  heating of the lowest axial mode at the rate of 180 phonons/ms and dephasing during two-qubit gates. In our experiments, the effect of crosstalk is negligible on logical gate performance.  

The axial heating can be reduced by cooling the trap \cite{LabaziewiczPRL2008}, preparing the trap surface \cite{HitePRL2012}, or both \cite{SedlacekPRA2018} and its effects can be countered by sympathetic cooling \cite{CetinaPRXQuantum2022}. Reduction of axial temperature together with a decrease in the axial phase gradients of the addressing beams should also reduce the two-qubit dephasing \cite{Sutherland2022dephasing}. Dephasing can be further reduced by suppressing the relative intensity fluctuations of the red and blue sidebands in a (fiber-coupled) double-pass AOM setup.

A limitation in our experiment is that detecting error syndromes does not help the decoding of the logical qubit without post-selection.  In all cases, it would be better not to apply the syndrome extraction gadget. Using the simulator, we explore a scenario where the heating rate is reduced to 0  and $Z$($X$) errors during two-qubit gates is reduced by a factor of $5(4)$.  We note that the axial motion is initially cooled to 5.5 mK, so the base uncertainty in the ion position remains due to residual 660 phonons on average.

 In this limit, we find that the logical error rate will drop to (0.8$\pm$ 0.06)\% for Steane error correction and (2.45$\pm$ 0.16)\% for Shor error correction (Table \ref{tab:EC}).   Both methods see substantial improvements in errors and Steane-error correction continues to outperform the Shor-error correction.  The post-selected states will be prepared with an improved rejection rate. For Steane with $\ket{0_L}$ ancilla, we see a reduction of rejection rates by a factor of 3 and error rates by a factor of 4 over the current simulated values (Table \ref{tab:postselection}).

\section{Conclusions}

Our work has demonstrated the promise of Steane-style quantum error correction. The next level of investigation requires intermediate measurements to measure and reset the Steane ancilla.  A larger system capable of pairing up multiple ion chains would allow us to perform Knill style syndrome extraction via teleportation.  Knill's method is both efficient and prevents the unwanted accumulation of leakeage errors to states outside of the qubit space. Leakage errors present a challenge to Shor and Steane-style error correction without additional leakage reduction gadgets \cite{AliferisQIC2007}. Further work should focus on examining codes with larger distance where the challenge of preparing logical state increases, but the gain from error correction also grows.  

The procedure outlined here is ideal for quantum architectures that enable low-error rate shuttling to connect distant qubits without multiple swap operations.  For ions, a CCD type architecture \cite{KielpinskiQCCD2002} that uses longer than two-ions chains as a baseline would work well. Ion temperature can be lowered after merging chains using sympathetic cooling and benefits from precooling one of the two merged chains. For neutral atoms, a change in encoding procedures from the Steane code \cite{bluvstein2023logical} would enable testing these results in that system.  In future quantum dot systems which allow coherent transport of spins, a similar result will hold \cite{TaylorNatPhys2005}. Steane error correction is also promising for a multi-layer architecture where a logical ancilla qubit layer can be directly coupled to the data qubit layer. 

\subsection*{Acknowledgements}
This work was funded by the Office of the Director of National Intelligence - Intelligence Advanced Research Projects Activity through an Army Research Office contract (W911NF-16-1-0082), the Army Research Office (W911NF-21-1-0005), the NSF STAQ project (Phy-1818914, Phy-2325080), the DOE QSA program, and the NSF QLCI for Robust Quantum Simulation (OMA-2120757). We thank L. Feng for help in building up the experimental setup at Duke University and Y. Yu for helpful discussions.

\subsection*{Related work} Please also note the preprint titled ``Demonstration of fault-tolerant Steane quantum error correction'' by the research groups led by M. M\"uller (Aachen \& J\"ulich) and by T. Monz (Innsbruck) on related work.

\subsection*{Author contributions} M.C. collected and analyzed the experimental data. S.H. analyzed the data and developed the numerical simulation. All authors developed the experimental protocol, discussed results, designed figures, and wrote the manuscript. 

\section*{Competing Interests}
K.R.B. is a scientific advisor for IonQ, Inc. and has a personal financial interest in the company. M.C. is a co-inventor on patents that are licensed by University of Maryland to IonQ, Inc.

\section*{\huge Supplementary Material}
\subsection*{Experimental System}

\begin{figure}
    \centering    
    \includegraphics[width=.9\linewidth]{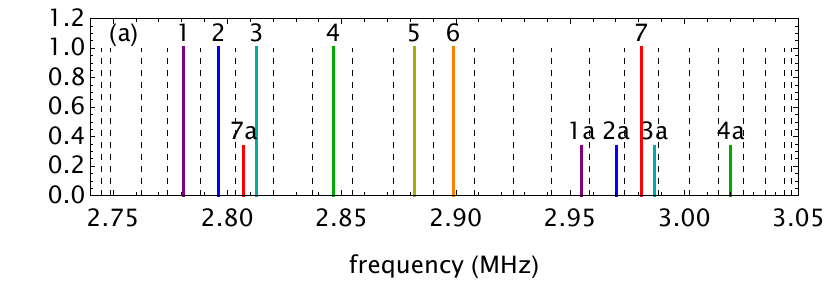}
    \caption{The frequencies of the radial modes of 23 ions (dashed lines) compared to the mean frequencies of the state-dependent amplitude-modulated forces (lines 1-7) that are used to perform entangling gates. Due to motion of the lowest axial mode of the ion chain, the gate waveforms as seen by the ions acquire sidebands at $\pm$174 kHz (lines 1a-7a).}
    \label{fig:mode_spectrum}
\end{figure}

To perform two-qubit M\o lmer-S\o rensen gates, we use individually-addressed Raman beams to induce state-dependent amplitude-modulated forces that couple the internal states of the ions to one set of the ions' radial modes (Fig. \ref{fig:mode_spectrum}). For robustness to mode frequency drift, we choose the mean frequencies of these forces (lines 1-7 on Fig. \ref{fig:mode_spectrum}) half-way between the frequencies of the neighbouring radial modes (dashed lines on Fig. \ref{fig:mode_spectrum}) and employ a time-reversal symmetric modulation waveforms that null the mean motional displacement of each mode during the gate \cite{KangPRAp2023}.

Predominant axial motion of the ions at the lowest axial mode frequency of 174 kHz causes the gate waveforms experienced by the ions to acquire modulation sidebands (lines 1a-7a on Fig. \ref{fig:mode_spectrum}). To suppress the resulting errors, we choose the XX gate waveforms for which these sidebands are separated from the radial modes by at least 3.3 kHz.

To mitigate the impact of the axial ion motion on the single-qubit gates, we first align the addressing beams to minimize the phase gradient of the Raman coupling along the chain axis. 
The effects of Raman amplitude fluctuations are countered by using SK1 pulses \cite{brown2004arbitrarily}. Each 13-$\mu$s long pulse segment of the compound pulse is further Gaussian-shaped so as to null the first-order contribution of the ions' motion at the lowest axial frequency to the resulting unitary.

\subsection*{Physical Error Rates and Error Modelling}

The logical error rate for each experiment is numerically estimated via Monte-Carlo sampling of state vector evolutions. 

The leading source of errors in the experiment is the motion of the ion chain in the  lowest axial mode (mode 0) with frequency $\omega_0 = 2\pi\times 174$ kHz, 
which misaligns the ions relative to the individual-addressing beams. These beams are nearly Gaussian-shaped with waist $w=450$ nm, corresponding to instantaneous carrier Rabi frequency of $i$-th ion $\Omega_{i} = \Omega_{i,\rm{max}} \exp(-x_i^2/w^2)$, where $x_i$ is the instantaneous position of this ion. 

Since, in our experiments, the energy $E_0$ of mode 0 corresponds to hundreds of  motional quanta $\hslash \omega_0$, we describe this motion classically. In this limit, $\Omega_{i}$ varies with the classical phase $\phi$ of the ions' motion as $\Omega_i = \Omega_{i,\rm{max}}\exp(- 2 \epsilon_i  \sin^2\phi)$, where $\epsilon_i = b_{i0}^2 \hslash n E_0 / m w^2 \omega_0^2$ is a unitless decay parameter, $m$ is the ion mass and $b_{i0}$ the participation of ion $i$ in mode $0$. Averaging over the motional phase $\phi$ gives the mean Rabi frequency $\bar{\Omega}_{i} = \Omega_{i,\rm{max}} f(\epsilon_i)$ of ion $i$, where $f(x) = e^{-x} I_0(x)$, with $I_0$ the modified Bessel function of the first kind.

Uncertainty in the energy $E_0$ introduces rotation errors in both single- and two-qubit gates \cite{CetinaPRXQuantum2022}. A carrier rotation of ion $i$ by angle $\theta$ becomes an imperfect rotation by angle $\theta f(\epsilon_i)$, while a two-qubit $XX(\theta)$ between ions $i$ and $j$ gate becomes $XX\left(\theta f(\epsilon_i)f(\epsilon_j)\right)$. Note that, in our model, $\epsilon_i$ and $\epsilon_j$ are perfectly correlated since they are both derived from the same random number $n$ of motional quanta as $\epsilon_{i,j}=n u_{i,j}$, with $u_{i,j} = b_{(i,j)0}^2 \hslash/(m w^2\omega_0)$ fixed unitless numbers that we calculate.

\begin{figure}
    \begin{center}
    \includegraphics[scale=0.7]{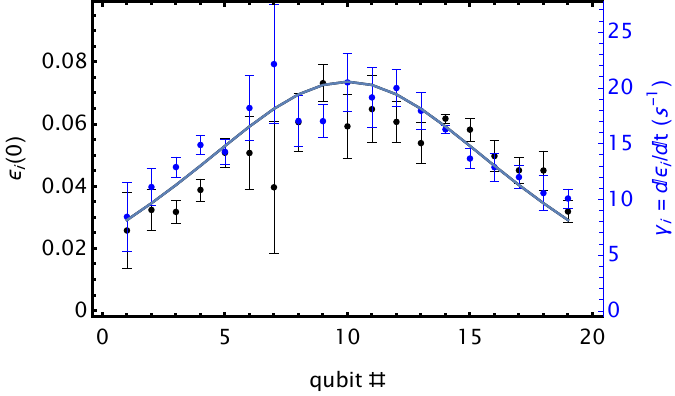}
    \caption{The decay parameter\cite{CetinaPRXQuantum2022} $\epsilon_i$ of carrier Rabi oscillations as function of qubit number at the beginning of the circuit (black) and the rate of its increase with time $\gamma_i$ (blue). Fits of the data to $\epsilon_i,\gamma_i\sim b_{i0}^2 $ (solid line) yield the initial mean phonon number in the lowest axial mode $\bar{n}=660(20)$ and the heating rate of the lowest axial mode of the chain of $\dot{\bar{n}}=$180(4) phonons/ms.}\label{fig:axial}
    \end{center}
\end{figure}

\begin{figure*}
    \begin{center}
    \begin{subfigure}
    \centering
    \includegraphics[width=.45\linewidth]{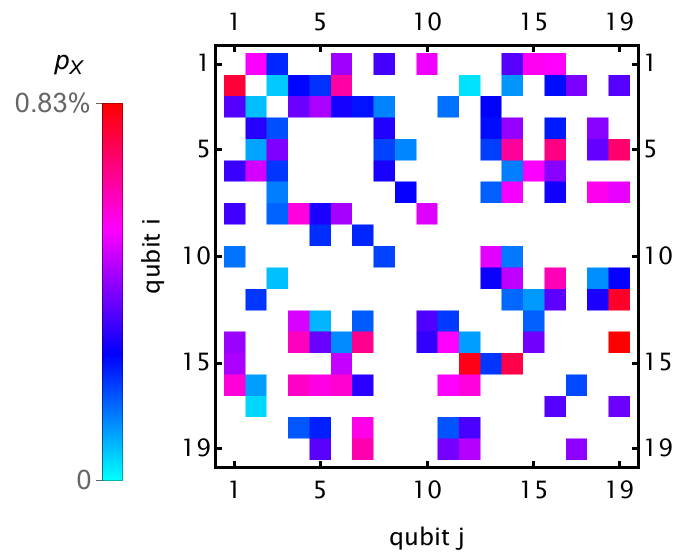}    
    \end{subfigure}
    \begin{subfigure}
    \centering
    \includegraphics[width=.45\linewidth]{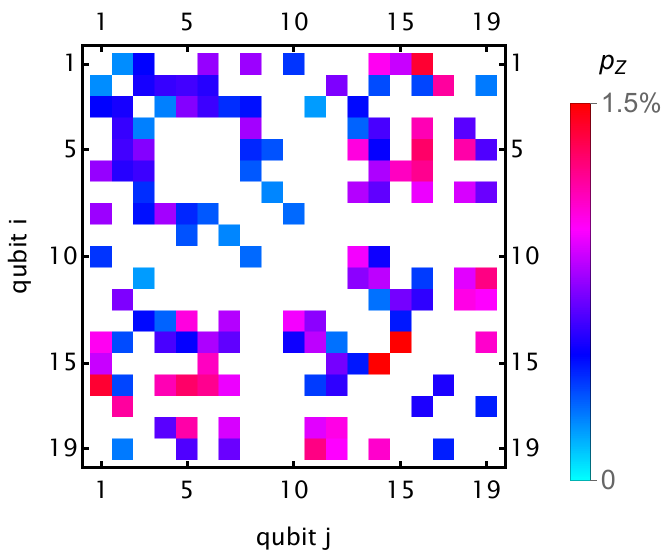}    
    \end{subfigure}   
    \caption{The measured probabilities $p_X$ of an $X$ error (a) and $p_Z$ of a $Z$ error (b), per qubit, per fully-entangling $XX$ gate between qubits $i$ and $j$.}\label{fig:XX_errors}
    \end{center}
\end{figure*}

We assume that at beginning of the circuit ($t=0$) mode $0$ is in thermal equilibrium. In this case, $\epsilon_i$ will be Boltzmann-distributed with mean $\bar{\epsilon_i}(0)$, which we determine by fitting carrier Rabi oscillations of all the qubits to the model from \cite{CetinaPRXQuantum2022} (Figure \ref{fig:axial}). 
We fit $\bar{\epsilon}_i$ to $\bar{n} u_i$ to obtain the mean initial phonon number $\bar{n}(0) = 660(20)$. At the beginning of the simulation, we sample $n$ from from the Boltzmann distribution with mean $\bar{n}(0)$. 

Without mid-circuit sympathetic cooling, the energy of the lowest axial mode increases with time, causing the gates applied near the end of the circuit to perform worse. We determine the rate of increase $\gamma_i$ of $\epsilon_i$ by fitting Rabi oscillations at later times in the circuit (Figure \ref{fig:axial}). A fit of $\gamma_i$ to $\dot{\bar{n}} \times u_i$ yields the axial heating rate $\dot{n}=$ 180(4) phonons/ms. This is consistent with a noisy electric field that is homogenous across the chain and corresponds to the heating rate of 8.3(2) phonons/ms of a single ion with axial frequency of $2\pi\times$ 174 kHz.

We model heating of mode 0 as a biased random walk in $n$. Given $n(r\Delta t)$, we set
\begin{eqnarray*}
   n((r+1)\Delta t) := \left\{\begin{array}{ll}
  n(r\Delta t) + 1& \textrm{with prob.\ } (n+1)\times\dot{\bar{n}}\Delta t\\
  n(r\Delta t) - 1& \textrm{with prob.\ } n\times\dot{\bar{n}}\Delta t\\
  n(r\Delta t) & \textrm{otherwise}.
   \end{array}\right.
\end{eqnarray*}

To determine the imperfect rotation angle for a gate starting at time $t$, for each raw single-qubit rotation $\sigma_\phi(\theta)$ in the $x$-$y$ plane on ion $i$, the actual rotation angle $\tilde{\theta}$ is set to be $$\tilde{\theta} := \theta \left(1+u_i\overline{n}(0)\right) f(u_i n),$$
The extra coefficient $(1+u_i \overline{n}(0))$ compensates the average under-rotation at $t=0$. Similarly, for each two-qubit $XX(\theta)$ rotation between ions $i$ and $j$ starting at time $t$, the actual rotation angle $\tilde{\theta}$ is set to be
$$\tilde{\theta} := \theta \left(1+u_i\overline{n}(0)\right) \left(1+u_j\overline{n}(0)\right) f(u_i n) f(u_j n).$$
We ignore  the fluctuation of $n(t)$ within a single- or two-qubit gate duration.

 Another contributing factor to logical error is dephasing during the entangling gates. Sources of this dephasing include axial phase gradients in the individual addressing beams combined with axial motion of the ions, fluctuations in the ratio of the red and blue sidebands and fluctuations in coupling to other Zeeman levels. We model the dephasing during the XX gate between qubits $i$ and $j$ as a pair of equal single-qubit phase-flip Pauli channels 
\begin{eqnarray}
\mathcal{E}_Z(\rho) = \left(1-p_Z^{(i,j)}\right) \rho + p_Z^{(i,j)} Z\rho Z
\end{eqnarray}
on qubits $i$ and $j$ following the gate.
We determine the error probability $p_Z^{(i,j)}$ for each ion pair by performing $K$ paired maximally entangling gates with opposite rotation angles ($+XX/-XX$) inside a Ramsey sequence on both ions. We fit the resulting Ramsey fringes to extract the contrast for both ions and set $p_Z^{(i,j)}$ to $1/2$ of the average measured contrast loss per gate (Fig. \ref{fig:XX_errors}). We assume $p_Z^{(i,j)}$ to be time-independent. 

To model dephasing on idling 
qubits due to optical phase fluctuations and unwanted AC shifts, for each idling circuit location with duration $t$, we apply a Pauli $Z$ error with probability $\Gamma_{\textrm{deph}} t$, where $\Gamma_{\textrm{deph}}$ is the measured dephasing rate.
For each measurement, we flip the outcome from $\ket{1}$ ($\ket{0}$) to $\ket{0}$ ($\ket{1}$) with probability $p_{1\rightarrow 0} = 4\times 10^{-3}$ ($p_{0\rightarrow 1} = 1.5\times 10^{-3}$).

\begin{figure}
    \begin{center}
    \includegraphics[width=.9\linewidth]{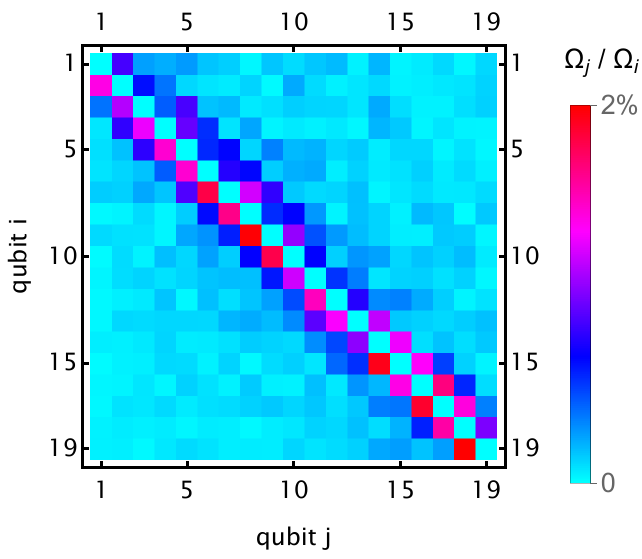}    
    \caption{The measured ratio of the Rabi frequency $\Omega_j$ of qubit $j$ and the 
    Rabi frequency $\Omega_i$ of the target resonantly driven qubit $i$.}
    \label{fig:crosstalk}
    \end{center}    
\end{figure}

Acoustic and electrical crosstalk in the 32-channel acousto-optic modulator (AOM) leads to crosstalk in individual addressing. We characterize this crosstalk by measuring the ratio of the carrier Rabi frequencies for a single addressed ion and the remaining qubits (Fig. \ref{fig:crosstalk}). The effects of crosstalk on single-qubit gates are suppressed using SK1 pulses. 

We estimate the contribution of two-qubit crosstalk errors on logical infidelity.
When a two-qubit gate $X_iX_j(\theta)$ is applied between ions $i$ and $j$ with carrier Rabi frequency $\Omega(t)$, each other ion $k \ne i,j$ will also experience a spin-dependent force with carrier Rabi frequency 
$\Omega_k(t) = \left(\chi_{i,k} + \chi_{j,k}\right) \Omega(t)$, where $\chi_{i,k}$ are the carrier crosstalk ratios. After the two-qubit drive is finished, two small two-qubit rotations $X_iX_k(\theta_{k}^{(1)})$ and $X_jX_k(\theta_{k}^{(2)})$ will be applied between ion pairs $(i,k)$ and $(j,k)$. 
Let $A_{i,j,k}$ be the ratio of rotation angle on ion pair $(i,k)$ when the two-qubit AM pulse sequence between $(i,j)$ is applied between $(i,k)$ instead. Then we have $$\theta_{k}^{(1)} = (\chi_{i,k}+\chi_{j,k}) A_{i,j,k}$$ and 
$$\theta_{k}^{(2)} = (\chi_{i,k}+\chi_{j,k}) A_{j,i,k}.$$
The values $A_{i,j,k}$ can be analytically calculated when the pulse shapes, motional mode frequencies and spin-motion coupling 

Motional dephasing during two-qubit gates causes $X$-errors \cite{Wang2020hifi} and 
contributes to the logical error rate. We determine the probability $p_X$ of single-qubit Pauli $X$ errors during the $X_iX_j(\theta)$ gate between ions $i$ and $j$ by preparing the target qubits in the $|0\rangle$ state and measuring the probability $p$ of a single bit-flip after $K$ fully entangling gates. We set $p_X = p/K/2$ and show the results in Fig. \ref{fig:XX_errors}b.

\subsection*{Fault-tolerant protocols}

\begin{table*}
    \centering
    \begin{tabular}{|c|c|c|c|c|c|c|} \hline 
          \multicolumn{2}{|c|}{\multirow{2}{*}{EC Protocol}} & \multicolumn{3}{c|}{Shor}& \multicolumn{2}{c|}{Steane}\\\cline{3-7} \multicolumn{2}{|c|}{}& Single shot&  Adaptive I&  Adaptive II&  $\ket{+_L}$& $\ket{0_L}$\\ \hline 
          \multirow{3}{*}{LER} &EXP &  $(9.72\pm 1.16)\%$  &  $(10.04\pm 1.45)\%$&  $(7.84\pm 1.57)\%$& $4.93_{-0.48}^{+0.51}\%$   & $4.75_{-0.47}^{+0.51}\%$ \\ \cline{2-7} 
          &SIM &  $(7.87\pm 0.17)\%$ &  $(7.92\pm 0.26)\%$&  $(6.11\pm 0.26)\%$&  $(4.25 \pm 0.13)\%$& $(4.15 \pm 0.12)\%$\\ \cline{2-7}
  &IMP & $(5.68\pm 0.14)\%$ & $(3.11\pm 0.17)\%$ & $(2.45\pm 0.16)\%$ & $(0.81 \pm 0.06)\%$& $(1.08 \pm 0.07)\%$\\\hline\hline
  \multirow{3}{*}{DSTB} & EXP & \multicolumn{3}{c|}{$(4.71\pm 1.59)\%$} & $1.45_{-0.26}^{+0.30}\%$ & $1.12_{-0.23}^{+0.26}\%$\\\cline{2-7}
   & SIM & \multicolumn{3}{c|}{$(2.41\pm0.10)\%$} & 
   $(1.16\pm 0.07)\%$& $(1.02\pm 0.06)\%$ \\\cline{2-7}
   & IMP & \multicolumn{3}{c|}{$(1.49\pm 0.08)\%$} & $(0.18\pm 0.03)\%$& $(0.21\pm 0.03)\%$ \\\hline
  
    \end{tabular}
    \caption{Comparison between Shor and Steane error correction protocols. We calculate the logical error rate (LER) and disturbance (DSTB) based on the experimental data (EXP), our simulation of the experiment (SIM), and our predicted simulation for future experimental improvements (IMP). 
    DSTB is the logical error rate when no ancilla readout information is used.
    Details of the protocols are in the main text.}
    \label{tab:EC}
\end{table*}

 \begin{table*}
    \centering
    \begin{tabular}{|c|c|c|c|c|c|} \hline 
         \multicolumn{2}{|c|}{PS Protocol}&  Direct $\ket{0_L}$ preparation&  Shor  &  Steane ($\ket{+_L}$ ancilla) & Steane  ($\ket{0_L}$ ancilla)\\ \hline 
         \multirow{3}{*}{LER} &EXP&  $0.20_{-0.14}^{+0.27}\%$ &  $1.30_{-0.49}^{+0.68}\%$ &  $0.66_{-0.21}^{+0.27}\%$& $0.34_{-0.14}^{+0.21}\%$\\ \cline{2-6} 
          &SIM& $0.22_{-0.03}^{+0.03}\%$ &  $0.82_{-0.07}^{+0.07}\%$ &   $0.46_{-0.05}^{+0.05}\%$& $0.26_{-0.04}^{+0.04}\%$\\ \cline{2-6} 
          &IMP& $0.14_{-0.02}^{+0.03}\%$ & $0.42_{-0.04}^{+0.05}\%$& $0.06_{- 0.02}^{+0.02}\%$& $0.06_{-0.01}^{+0.02}\%$\\ \hline 
          \multirow{3}{*}{RR}&EXP& \multirow{3}{*}{N/A} &  $(35.32 \pm 1.87) \%$ &  $(27.10 \pm 1.44)\%$ & $(33.81 \pm 1.32)$\% \\ \cline{2-2}\cline{4-6}
          &SIM&  &  $(33.04\pm 0.29)\%$&  $(33.67 \pm 0.36)\%$& $(32.16 \pm 0.35) \%$\\ \cline{2-2} \cline{4-6}
          &IMP&  & $(20.71\pm 0.25)\%$ & $(15.98 \pm 0.25)\%$ & $(12.32 \pm 0.23)\%$\\ \hline
    \end{tabular}
    \caption{Comparison of postselection protocols for $\ket{0_L}$ between Shor- and Steane-style error detection ciruits. We calculate the logical error rate (LER) based on the experimental data (EXP), our simulation of the experiment (SIM), and our predicted simulation for future experimental improvements (IMP). Postselection leads to a rejection rate (RR) which is the fraction of trials that are not postselected.}
    \label{tab:postselection}
\end{table*}

Here we describe the  fault-tolerant protocols implemented in our experiments, which include circuits for logical state preparation and measurement (SPAM) and syndrome extraction, decoders and calculation of logical error rates (LER). We focus on how syndrome extraction circuits disturb the logical $\ket{0_L}$ state of the $[[9,1,3]]$ Bacon-Shor code and only consider logical bit-flip ($X_L$) errors when calculating LER. 
We compare  experiment 
results with
numerical simulations to benchmark our error modeling, and provide predictions of LER assuming improved experimental parameters.
In our prediction, we assume that sympathetic cooling will cool down the lowest axial mode to initial temperature before each single- and two-qubit gate is performed, which resets the variable $\epsilon$ by a random number from the Boltzmann distribution.
We also assume a $5\times$ reduction of $Z$ errors during the two-qubit gates and a $4\times$ reduction of $X$ errors from imperfect spin-motion disentanglement. For all the result tables,  we use LER and RR to denote logical error rate and rejection rate of postselection, and 
EXP, SIM, IMP to denote rates extracted from experimental data, simulations with measured experimental parameters during the preparation of this work and simulation with anticipated improvement of experimental parameters. The main experimental and numerical results are presented in Tables \ref{tab:EC} and \ref{tab:postselection}.

\subsubsection*{Logical state preparation and measurement}
The subsystem $[[9,1,3]]$ Bacon-Shor code has four stabilizer generators $Z_1Z_2Z_3Z_4Z_5Z_6$, $Z_4Z_5Z_6Z_7Z_8Z_9$, $X_1X_2X_4X_5X_7X_8$ and $X_2X_3X_5X_6X_8X_9$. The gauge operators are $Z_iZ_{i+3}$ ($i=1,\ldots,6$) and $X_iX_{i+1}$ ($i\in\{1,2,4,5,7,8\}$).

\begin{figure*}
    \includegraphics[width=0.45\linewidth]{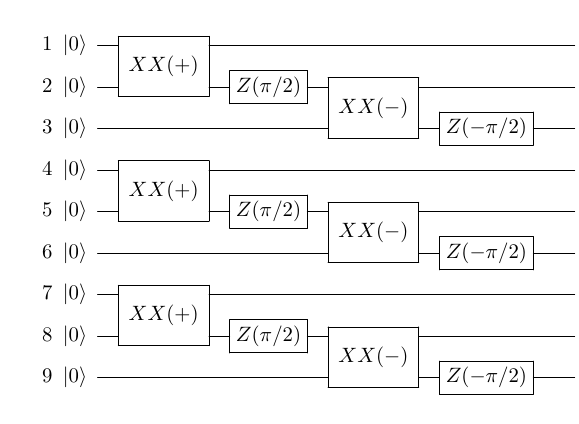}
    \includegraphics[width=0.525\linewidth]{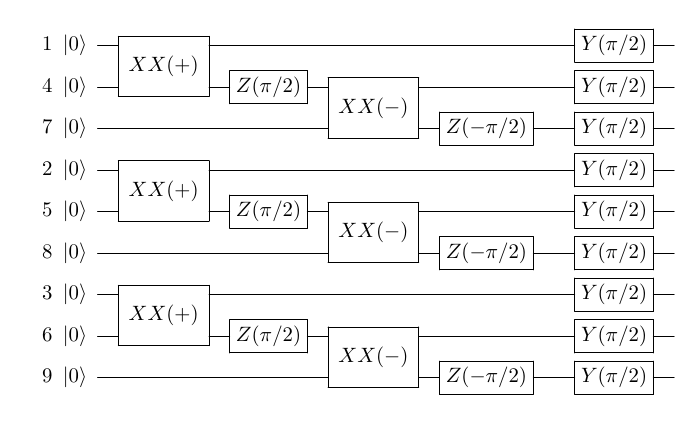}
       
    \caption{Left (right): the logical $\ket{0}$ ($\ket{+}$) state preparation circuit using trapped-ion native operations. $XX(\pm) = XX(\pm\pi/4)$.}\label{fig:prep}
\end{figure*}

The logical $\ket{0_L}$ ($\ket{+_L}$) state consists of three rows (columns) of physical $\ket{+++}+\ket{---}$ ($\ket{000}+\ket{111}$) GHZ 
states, which can be prepared fault-tolerantly with unitary operations only~\cite{li2018direct}. 
 The implementations for $\ket{0_L}$ and $\ket{+_L}$ state preparations using trapped-ion native operations are shown in Figure ~\ref{fig:prep}.
In these circuits, any single- or two-qubit error of weight $1$ is equivalent to a single error on the code block up to gauges.

The logical operator $Z_L = Z_1Z_2Z_3$ 
 can be fault-tolerantly measured
by transversal $Z$ readouts followed by a majority vote among the row parities $Z_{3i+1}Z_{3i+2}Z_{3i+3}$ ($i=0,1,2$).
Similarly, $X_L = X_1X_4X_7$ can be measured by transversal $X$ readouts followed by a majority vote among the column parities 
$X_{1+j}Z_{4+j}Z_{7+j}$ ($j=0,1,2$).

\begin{figure*}
    \includegraphics[width=\linewidth]{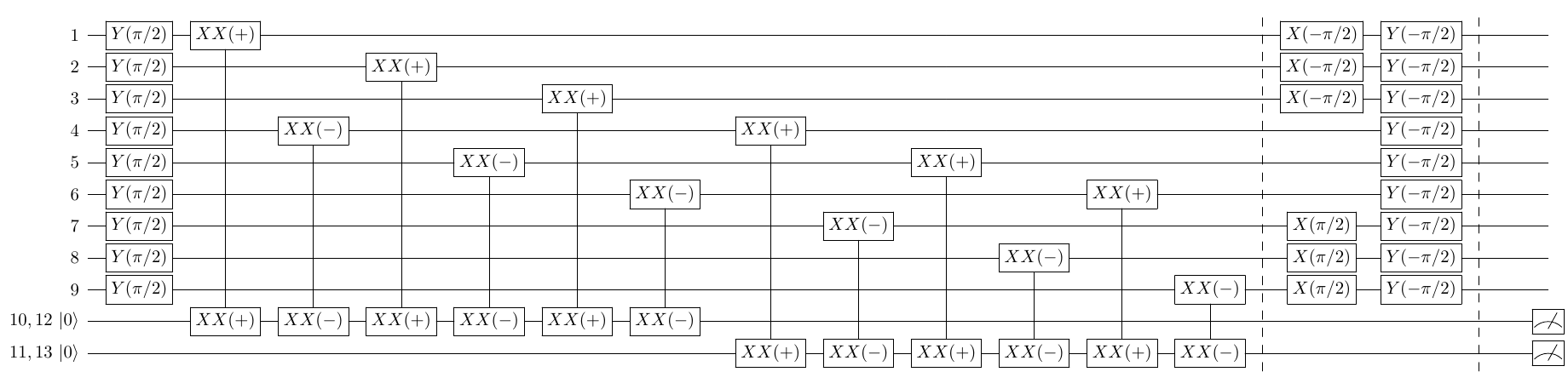}
    \caption{Implementation of Shor-style syndrome extraction using trapped-ion native gates. $XX(\pm) = XX(\pm\pi/4)$. }
    \label{fig:circuit_shor}
\end{figure*}

\subsubsection*{Shor-style syndrome extraction}

\begin{table*}[]
    \centering
    \begin{tabular}{|c|c|c|c|c|c|}
    \hline
        \multicolumn{2}{|c|}{$s^{(1)}$}  & 00 & 10 & 11 & 01\\ \cline{1-6}
        \multirow{3}{*}{$\mu_1$} & EXP & $64.68_{-1.91}^{+1.88}\%$&$10.40_{-1.17}^{+1.26}\%$&$7.20_{-0.98}^{+1.08}\%$&$17.72_{-1.48}^{+1.55}\%$ \\\cline{2-6}
         & SIM & $66.96_{-0.29}^{+0.29}\%$&$11.27_{-0.20}^{+0.20}\%$&$7.04_{-0.16}^{+0.16}\%$&$14.73_{-0.22}^{+0.22}\%$\\\cline{2-6}
         & IMP  & $79.29_{-0.25}^{+0.25}\%$&$6.44_{-0.15}^{+0.15}\%$&$3.87_{-0.12}^{+0.12}\%$&$10.40_{-0.19}^{+0.19}\%$\\\hline
        \multirow{3}{*}{$\lambda_1$} & EXP & $1.30_{-0.49}^{+0.68}\%$&$14.23_{-4.01}^{+4.85}\%$&$16.11_{-5.05}^{+6.20}\%$&$35.21_{-4.45}^{+4.65}\%$\\\cline{2-6}
         & SIM & $0.82_{-0.07}^{+0.07}\%$&$10.71_{-0.57}^{+0.59}\%$&$17.02_{-0.87}^{+0.90}\%$&$33.38_{-0.76}^{+0.77}\%$\\
        \cline{2-6}
        & IMP & $0.42_{-0.04}^{+0.05}\%$&$5.96_{-0.57}^{+0.61}\%$&$8.24_{-0.85}^{+0.91}\%$&$44.68_{-0.96}^{+0.96}\%$\\\hline
         \multirow{3}{*}{$\delta_1$} & EXP & $1.30_{-0.49}^{+0.68}\%$&$6.92_{-2.77}^{+3.80}\%$&$20.00_{-5.58}^{+6.60}\%$&$9.71_{-2.59}^{+3.15}\%$\\ \cline{2-6}
          & SIM & $0.82_{-0.07}^{+0.07}\%$&$3.91_{-0.35}^{+0.37}\%$&$9.31_{-0.67}^{+0.70}\%$&$5.23_{-0.35}^{+0.37}\%$\\\cline{2-6}
& IMP &$0.42_{-0.04}^{+0.05}\%$&$3.43_{-0.43}^{+0.47}\%$&$9.64_{-0.91}^{+0.97}\%$&$5.44_{-0.43}^{+0.45}\%$\\\hline\hline
        \multirow{3}{*}{$\mu_2$} & EXP & $64.01_{-1.10}^{+1.09}\%$&$10.85_{-0.70}^{+0.73}\%$&$8.33_{-0.62}^{+0.65}\%$&$16.80_{-0.84}^{+0.87}\%$\\\cline{2-6}
         & SIM & $67.01_{-0.29}^{+0.29}\%$&$10.95_{-0.19}^{+0.20}\%$&$6.92_{-0.16}^{+0.16}\%$&$15.12_{-0.22}^{+0.22}\%$\\\cline{2-6}
         & IMP & $79.36_{-0.25}^{+0.25}\%$&$6.53_{-0.15}^{+0.15}\%$&$3.79_{-0.12}^{+0.12}\%$&$10.32_{-0.19}^{+0.19}\%$\\\hline
       \multirow{3}{*}{$\lambda_2$} & EXP & $13.31_{-0.95}^{+0.99}\%$&$25.18_{-2.95}^{+3.13}\%$&$30.08_{-3.57}^{+3.76}\%$&$24.60_{-2.36}^{+2.48}\%$\\\cline{2-6}
        & SIM & $10.07_{-0.23}^{+0.23}\%$&$18.64_{-0.73}^{+0.74}\%$&$30.04_{-1.08}^{+1.10}\%$&$21.45_{-0.65}^{+0.66}\%$\\
       \cline{2-6}
       & IMP & $6.14_{-0.17}^{+0.17}\%$&$9.69_{-0.71}^{+0.74}\%$&$19.46_{-1.25}^{+1.30}\%$&$13.60_{-0.66}^{+0.68}\%$\\\hline
        \multirow{3}{*}{$\delta_2$} & EXP & $8.14_{-0.76}^{+0.81}\%$&$19.04_{-2.64}^{+2.87}\%$&$26.56_{-3.42}^{+3.65}\%$&$18.02_{-2.09}^{+2.24}\%$\\\cline{2-6}
         & SIM & $4.49_{-0.16}^{+0.16}\%$&$11.66_{-0.60}^{+0.62}\%$&$19.78_{-0.93}^{+0.96}\%$&$13.98_{-0.55}^{+0.56}\%$\\
         \cline{2-6}
         & IMP &$2.40_{-0.11}^{+0.11}\%$&$8.16_{-0.65}^{+0.69}\%$&$17.74_{-1.20}^{+1.25}\%$&$11.59_{-0.61}^{+0.63}\%$\\\hline
    \end{tabular}
    \caption{The probability table for Shor-style syndrome extraction conditioned on the syndrome $s^{(1)}$ in the first round.  We calculate the probability of observing the first syndrome $\mu_k$, the logical error rate $\lambda_k$ with ancilla based correction, and the disturbance $\delta_k$ which is the logical error rate with only correcting on data measurement for $k=1$ or $k=2$ measurement of the stabilizer generators. We calculated these values based on the experimental data (EXP), our simulation of the experiment (SIM), and our predicted simulation for future experimental improvements (IMP). }
    \label{tab:Shor}
\end{table*}

Shor error correction measures each stabilizer element individually~\cite{ShorCatState}. 
We experimentally perform repetitive measurements  of two $Z$-type stabilizer elements $S_1 = Z_1Z_2Z_3Z_4Z_5Z_6$ and $S_2 = Z_4Z_5Z_6Z_7Z_8Z_9$ on $\ket{0_L}$ of $[[9,1,3]]$ Bacon-Shor code.

The measurement circuit for each syndrome extraction round using trapped-ion native gates is shown in Figure \ref{fig:circuit_shor}, which only utilizes two ancilla qubits.
Our choice of two-qubit gate scheduling guarantees that any single fault an ancilla only propagate one error to data up to gauges~\cite{li2018direct}.

For a distance-$3$ code, at most $2$ round of extractions is sufficient for fault tolerance~\cite{chao2018quantum}. 
If the decoder cannot determine the correction from the syndrome in the first round,
a second extraction round will be triggered.
We decode the experimental data with three different decoders. For convenience, we denote the binary outcomes in round $i$ by a two-bit string $s^{(i)} = s_1^{(i)}s_2^{(i)} \in \{0,1\}^2$. The three decoders are described as follows:

(i) \textbf{Single-shot decoder:} Only perform a single extraction round and apply correction with $s^{(1)}$ immediately. 
The corresponding correction for $s^{(1)} = 00$, $01$, $11$ and $10$ are $I$, $X_1$, $X_4$ and $X_7$, respectively. 
This decoder is not fault-tolerant since a single $X$ error in between two CNOT gates on qubits 4-6 will have syndrome $s^{(1)} = 01$.
After correction, there will be two $X$ errors on the second and third qubit rows, which are uncorrectable.
In our experiment, these harmful errors are $Z$ errors between two $XX$ gates on qubits 4-6. The single-qubit gates after $XX$ gates will rotate these $Z$ errors into $X$ errors.

(ii) \textbf{Adaptive decoder I}: 
If $s^{(1)} = 00$, we do not perform the next extraction round and no correction will be applied. 
Otherwise, the next extraction round will be triggered. As $s^{(1)}$ is non-trivial, we can assume that $s^{(2)}$ is trustworthy. The corresponding correction for $s^{(2)} = 00$, $01$, $11$ and $10$ are $I$, $X_1$, $X_2$ and $X_3$, respectively.

(iii) \textbf{Adaptive decoder II}:
The adaptive decoder I can be further improved 
by noticing that the second extraction round is unnecessary when $s^{(1)} = 10,11$.
When $s^{(1)} = 10$, the compatible weight-$1$ faults are either single-qubit $X$ error on qubits 1-3 before the two-qubit gates, or a single measurement error on $s_1^{(1)}$. 
If we apply a correction $X_1$ immediately, we either successfully correct the data qubit error (up to a gauge), or convert the measurement error into a data-qubit error without amplification of errors. 
If $s^{(1)} = 11$, the unique compatible weight-$1$ fault is a single-qubit $X$ error on qubits 4-6 before all two-qubit gates. Thus one can immediately apply a correction $X_2$.
If $s^{(1)} = 01$, another extraction round is performed and correction is applied based on $s^{(2)}$.  In fact, we only need to measure $S_1$ in the second round~\cite{delfosse2020short}, which can further minimize the disturbance to the data block. This optimization is not considered in our experiment.

Our current experimental setup does not support mid-circuit measurements and therefore cannot execute adaptive circuits. 
To estimate the logical error rate of adaptive decoders,
we post-select data from two experiments,
denoted by $E_1$ and $E_2$, in which single and two extraction rounds are deterministically performed.
For $E_2$, we do not refresh the ancilla but use another two ancilla qubits for the second extraction round.
Let $\mu_r(s^{(1)})$ ($r=1,2$) be the probability that $s^{(1)}$ is measured in the experiment $E_r$, 
$\lambda_{r}(s^{(1)})$ ($r=1,2$) be the conditional logical error rate when $E_r$ is performed with first-round syndrome $s^{(1)}$ while $s^{(r)}$ is used for correction. For single-shot decoder, The total logical error rates
for single-shot decoder, adaptive decoder I and adaptive decoder II are
\begin{eqnarray*}
p_{L,\textrm{ss}} &=& \mu_1(00)\lambda_1(00)+\mu_1(10)\lambda_1(10)\\&+&\mu_1(11)\lambda_1(11)+\mu_1(01)\lambda_1(01)\\
p_{L,\textrm{a1}} &=& \mu_1(00)\lambda_1(00)+
\mu_1(10)\lambda_2(10)\\
&+& \mu_1(11)\lambda_2(11)+\mu_1(01)\lambda_2(01)\\
p_{L,\textrm{a2}} &=& \mu_1(00)\lambda_1(00)+
\mu_1(10)\lambda_1(10)\\
&+& \mu_1(11)\lambda_1(11)+\mu_1(01)\lambda_2(01).
\end{eqnarray*}
Note that we have ignored the potential errors accumulated in future mid-circuit measurements. 

Table \ref{tab:Shor} shows the values of $\mu_r(s^{(1)})$ and $p_r(s^{{(1)}})$ in both experiments and simulations. 
We found that $\lambda_1(s^{(1)}) < \lambda_2(s^{(1)})$ for $s^{(1)} = 00, 10, 11$ while 
$\lambda_1(s^{(1)}) > \lambda_1(s^{(1)})$ for $s^{(1)} = 01$. This indicates that single-shot decoder is insufficient for fault tolerance.
We also see that the experimental values of $\mu_1(s^{(1)})$ and $\mu_2(s^{(1)})$ agree with negligible difference.

Table \ref{tab:Shor} also show the values of the \textit{conditional disturbance} $\delta_r(s^{(1)})$, defined as the logical error rate of $E_r$ conditioned on $s^{(1)}$ without the use of ancilla information $s^{(i)}$.
In most cases $\lambda_r(s^{(1)}) > \delta_r(s^{(1)})$, indicating that ancilla information is not helping. We found that simulation values of $\delta_r$ are all about $40\%$ smaller than experiment values. In our circuit, the majority of disturbance is contributed from dephasing errors in two-qubit gates. This mismatch 
might indicate that the dephasing errors in two-qubit gates can be time-dependent and correlated.

The experimental and theoretical values of logical error rates are presented in Table.~\ref{tab:EC}. We see that adaptive decoder II already outperforms single-shot decoder in the experiment.
Simulation shows that adaptive decoder I will also outperforms single-shot with improved error parameters, which further verifies that repetitive stabilizer measurements are necessary in Shor-style error correction. 

We also consider the use of one-round Shor-style circuit for postselection of $\ket{0_L}$, whose logical error and rejection rates are $\lambda_1(00)$ and $1-\mu_1(00)$. From Table \ref{tab:postselection} we see that Shor-style postselection introduces much more logical $X$ errors than direct preparation of $\ket{0_L}$ ($1.30_{-0.49}^{+0.68}\%$ vs. $0.20_{-0.14}^{+0.27}\%$), with the extra cost of non-vanishing rejection rate.

\subsubsection*{Steane-style syndrome extraction}
\begin{table*}[]
    \centering
    \begin{tabular}{|p{2.65cm}|c|c|c|c|c|}
    \hline
    Circuits & $\textrm{CNOT}_{L}\ket{0_L}\ket{+_L}$ & $\ket{0_L}\ket{+_L}$ & $\textrm{CNOT}_{L}\ket{0_L}\ket{0_L}$ &
     $\ket{0_L}\ket{0_L}$ & $\ket{0_L}$ only\\ \hline
   LER (without ancilla feedback) & $1.45_{-0.26}^{+0.30}\%$& $0.32_{-0.18}^{+0.31}\%$ & $
    1.12_{-0.37}^{+0.49}\%$& $0.40_{-0.21}^{+0.33}\%$ & $0.20_{-0.14}^{+0.27}\%$\\
   \hline
    LER  (with ancilla feedback)     & $4.93_{-0.48}^{+0.51}\%$ & \multirow{2}{*}{N/A}
    & $3.20_{-0.47}^{+0.53}\%$ & \multirow{2}{*}{N/A} 
    & \multirow{2}{*}{N/A}\\
    \cline{1-2}\cline{4-4}
    LER (with ancilla PS)     & $0.66_{-0.21}^{+0.27}\%$&  & $0.34_{-0.19}^{+0.33}\%$ &   & \\\cline{1-5}
    RR (ancilla PS)     & $(35.80 \pm 1.35)$\% & $(11.36 \pm 1.32) \%$ & $(27.10 \pm 1.44)\%$ & $(6.52 \pm 1.00)\%$ & \\\hline
    LER (data PS)  & $0.13_{-0.07}^{+0.12}\%$ & $0.04_{-0.04}^{+0.20}\%$ & $<0.09\%$ & $<0.16\%$ & $0.04_{-0.04}^{+0.19}\%$\\\hline
    RR (data PS)     & $(17.19 \pm 0.94)$\% & $(6.36 \pm 0.99)\%$ & $(14.44 \pm 1.05)\%$ & $(6.88 \pm 1.03)\%$ & $(5.08 \pm 0.88)\%$\\\hline
    LER (ancilla+data PS)     & $0.05_{-0.05}^{+0.21}\%$ & N/A & $<0.11\%$& N/A & \multirow{2}{*}{N/A}\\
    \cline{1-5}
    RR (ancilla+data PS)     & $(42.57 \pm 1.48)$\% & $(16.88 \pm 1.61)\%$ & $(32.68 \pm 1.58)\%$ & $(12.28 \pm 1.37)\%$ & \\\hline
    \end{tabular}
    \caption{Experiment result of Steane-error correction (EC) and postselection (PS). We consider two preparations of the logical ancilla qubit $\ket{A_L}$= $\ket{0_L}$ or $\ket{A_L}$=$\ket{+_L}$. Steane requires a logical CNOT and we compare experimental results when we apply it ($\textrm{CNOT}_{L}\ket{0_L}\ket{A_L}$) or not ($\ket{0_L}\ket{A_L}$). We also examine the experimental results when the ancilla qubit is not prepared ($\ket{0_L}$ only). 
    }
    \label{tab:SteaneExptCompare}
\end{table*}

\begin{figure}
    \centering
    \includegraphics{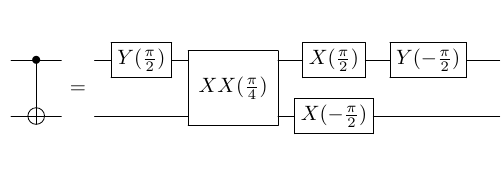}
    \caption{Physical CNOT implementation  using trapped-ion native operations. In a transversal CNOT gate circuit, the physical CNOT circuits are applied sequentially on each data-ancilla pair.}
    
    \label{fig:CNOT}
\end{figure}

In Steane error correction~\cite{SteaneSteaneEC1997},
a transversal CNOT gate is applied between data and a fault-tolerantly prepared $\ket{+_L}$ ($\ket{0_L}$) ancilla resource state used for correcting $X$ ($Z$) errors.
When correcting $X$ ($Z$) errors, the data (ancilla) block is the control of CNOT, while the ancilla (data) block is the target. After the transversal CNOT gate, all qubits are measured in $Z$- ($X$-) basis.
The error syndrome is obtained by parities of $Z$- ($X$-) stabilizer elements. 
We then assume the syndrome is perfect and
apply correction immediately based on the syndrome. In our Steane EC experiment, we first sequentially prepare a $\ket{0_L}$ state for data and a $\ket{+_L}$ state for ancilla, then apply transversal CNOT between data and ancilla, and finally measure all physical qubits in $Z$ basis. 
The physical implementation of each CNOT using trapped-ion operations is presented in Figure \ref{fig:CNOT}. The CNOT gates are applied sequentially in our experiment.

For benchmarking purposes, we also consider replacing $\ket{+_L}$ by $\ket{0_L}$ for $X$ error correction. 
Normally this does not work as it performs a logical $Z_L$ measurement on data. In this case it works because we are guaranteed the data should start in an eigenstate of $Z_L$. To show the errors induced by the transversal CNOT and to see the interaction between the data and the syndrome qubits, we compare to an experiment where we do not do the CNOT. 

The experimental results are summarized in Table \ref{tab:SteaneExptCompare}. Similar to Shor EC, feedback from ancilla introduces extra errors beyond the disturbance due to the EC circuit. We also postselect the experiment data with non-trivial (i) ancilla syndrome, (ii) data syndrome, or (iii) joint syndrome. We found that in the experiments with transversal CNOT applied, the rejection rate of (i) + (ii) exceeds that of (iii), indicating that transversal CNOT gate does propagate $X$ errors from data to ancilla. Note that (i) on $\textrm{CNOT}_L\ket{0_L}\ket{0_L}$ is exactly the Steane-style state-preparation protocol, in which fault tolerance of the initial $\ket{0_L}$ states are often not assumed. The $Z_L$ operator on ancilla can be served as an extra check for decreasing LER of postselected $\ket{0_L}$ state.
As our direct preparation of $\ket{0_L}$ is already fault tolerant, we do not expect that Steane state preparation will significantly outperform the unitary protocol. 
The simulation results for $\textrm{CNOT}_L \ket{0_L}\ket{+_L}$ and $\textrm{CNOT}_L\ket{0_L}\ket{0_L}$ can be found in Tables \ref{tab:EC} and \ref{tab:postselection}.

\subsubsection*{Logical Bell-state preparation}

We also study the performance of logical Bell state $\ket{0_L0_L}+\ket{1_L1_L}$, which can be directly prepared using the circuit $\textrm{CNOT}_L \ket{+_L}\ket{0_L}$. We use the same set of control and target qubits as in the Steane EC experiment (see Figure \ref{fig:circuitsandions}e). 
We measure all qubits in $Z$ ($X$) basis to extract the logical $ZZ$ ($XX$) outcome.
We first decode the measurement outcome of each logical qubit separately, then check whether the logical parity flips. We do not perform correlated decoding over two logical blocks.
For logical $X$ readout, 
transversal $Y(-\pi/2)$ rotations are performed on all qubits before the measurement, and the majority vote is performed over the parity of each column.
The experiment and simulation results are presented in Table \ref{tab:bell_state}. 
We find that logical $XX$ readout is significantly worse than $ZZ$ in experiment. For the optimistic error budget improvement we have considered, the improvement of logical $XX$ readout is also not as significant as $ZZ$.
We believe that the readout fidelity of $XX$ is limited by both under-rotation errors in the final single-qubit $Y$ gates and idle dephasing.
\begin{table}
    \centering
    \begin{tabular}{|c|l|c|c|} \hline 
          \multicolumn{2}{|c|}{Logical operator}&  $ Z_{L,1}Z_{L,2}$& $ X_{L,1}X_{L,2}$\\ \hline 
          \multirow{3}{*}{LER (EC)}&EXP&  $6.50_{-1.04}^{+1.17}\%$& $33.75_{-2.07}^{+2.12}\%$\\ \cline{2-4} 
          &SIM&  $(7.91 \pm 0.17)\%$& $(26.73 \pm 0.43)\%$\\ \cline{2-4} 
          &IMP&  $(1.25 \pm 0.07)\%$& $(18.96 \pm 0.24)\%$\\ \hline 
          \multirow{3}{*}{LER (PS)} &EXP&  $0.58_{-0.34}^{+0.61}\%$& $16.13_{-2.88}^{+3.22}\%$\\ \cline{2-4} 
          &SIM&  $(1.02 \pm 0.08)\%$& $(7.05 \pm 0.53)\%$ \\ \cline{2-4} 
          &IMP&  $< 10^{-4}$& $(2.01 \pm 0.16)\%$\\ \hline 
          \multirow{3}{*}{RR (PS)} &EXP&  $(39.15 \pm 2.74)\%$& $(70.55 \pm 3.68)\%$\\ \cline{2-4} 
          &SIM&  $(37.77 \pm 0.38)\%$& $(77.41 \pm 0.86)\%$ \\ \cline{2-4} 
          &IMP&  $(18.81 \pm 0.27)\%$& $(71.23 \pm 0.52)\%$\\ \hline
    \end{tabular}
    \caption{Experiment and simulation results for Bell-state preparation. A logical error occurs when the measured joint logical parity is $-1$.}
    \label{tab:bell_state}
\end{table}

\bibliography{References}

\begin{thebibliography}{10}
\urlstyle{rm}
\expandafter\ifx\csname url\endcsname\relax
  \def\url#1{\texttt{#1}}\fi
\expandafter\ifx\csname urlprefix\endcsname\relax\def\urlprefix{URL }\fi
\expandafter\ifx\csname doiprefix\endcsname\relax\def\doiprefix{DOI: }\fi
\providecommand{\bibinfo}[2]{#2}
\providecommand{\eprint}[2][]{\url{#2}}

\bibitem{GidneyQuantum2021}
\bibinfo{author}{Gidney, C.} \& \bibinfo{author}{Ekerå, M.}
\newblock \bibinfo{journal}{\bibinfo{title}{How to factor 2048 bit rsa integers in 8 hours using 20 million noisy qubits}}.
\newblock {\emph{\JournalTitle{Quantum}}} \textbf{\bibinfo{volume}{5}}, \bibinfo{pages}{433}, \doiprefix\url{10.22331/q-2021-04-15-433} (\bibinfo{year}{2021}).

\bibitem{microsoft2022}
\bibinfo{author}{Beverland, M.~E.} \emph{et~al.}
\newblock \bibinfo{journal}{\bibinfo{title}{Assessing requirements to scale to practical quantum advantage}}.
\newblock {\emph{\JournalTitle{arXiv:2211.07629}}}  (\bibinfo{year}{2022}).
\newblock \eprint{2211.07629}.

\bibitem{EganNature2021}
\bibinfo{author}{Egan, L.} \emph{et~al.}
\newblock \bibinfo{journal}{\bibinfo{title}{Fault-tolerant control of an error-corrected qubit}}.
\newblock {\emph{\JournalTitle{Nature}}} \textbf{\bibinfo{volume}{598}}, \bibinfo{pages}{281--286}, \doiprefix\url{10.1038/s41586-021-03928-y} (\bibinfo{year}{2021}).

\bibitem{NguyenPPAppl2021}
\bibinfo{author}{Nguyen, N.~H.} \emph{et~al.}
\newblock \bibinfo{journal}{\bibinfo{title}{Demonstration of shor encoding on a trapped-ion quantum computer}}.
\newblock {\emph{\JournalTitle{Phys. Rev. Appl.}}} \textbf{\bibinfo{volume}{16}}, \bibinfo{pages}{024057}, \doiprefix\url{10.1103/PhysRevApplied.16.024057} (\bibinfo{year}{2021}).

\bibitem{SivakNature2023}
\bibinfo{author}{Sivak, V.~V.} \emph{et~al.}
\newblock \bibinfo{journal}{\bibinfo{title}{Real-time quantum error correction beyond break-even}}.
\newblock {\emph{\JournalTitle{Nature}}} \textbf{\bibinfo{volume}{616}}, \bibinfo{pages}{50–55}, \doiprefix\url{10.1038/s41586-023-05782-6} (\bibinfo{year}{2023}).

\bibitem{RyanAndersonPRX2021}
\bibinfo{author}{Ryan-Anderson, C.} \emph{et~al.}
\newblock \bibinfo{journal}{\bibinfo{title}{Realization of real-time fault-tolerant quantum error correction}}.
\newblock {\emph{\JournalTitle{Phys. Rev. X}}} \textbf{\bibinfo{volume}{11}}, \bibinfo{pages}{041058}, \doiprefix\url{10.1103/PhysRevX.11.041058} (\bibinfo{year}{2021}).

\bibitem{RyanAndersonArxiv2022}
\bibinfo{author}{Ryan-Anderson, C.} \emph{et~al.}
\newblock \bibinfo{title}{Implementing fault-tolerant entangling gates on the five-qubit code and the color code} (\bibinfo{year}{2022}).

\bibitem{PostlerNature2022}
\bibinfo{author}{Postler, L.} \emph{et~al.}
\newblock \bibinfo{journal}{\bibinfo{title}{Demonstration of fault-tolerant universal quantum gate operations}}.
\newblock {\emph{\JournalTitle{Nature}}} \textbf{\bibinfo{volume}{605}}, \bibinfo{pages}{675–680}, \doiprefix\url{10.1038/s41586-022-04721-1} (\bibinfo{year}{2022}).

\bibitem{KrinnerNature2022}
\bibinfo{author}{Krinner, S.} \emph{et~al.}
\newblock \bibinfo{journal}{\bibinfo{title}{Realizing repeated quantum error correction in a distance-three surface code}}.
\newblock {\emph{\JournalTitle{Nature}}} \textbf{\bibinfo{volume}{605}}, \bibinfo{pages}{669--674}, \doiprefix\url{10.1038/s41586-022-04566-8} (\bibinfo{year}{2022}).

\bibitem{SundaresanNatComm2023}
\bibinfo{author}{Sundaresan, N.} \emph{et~al.}
\newblock \bibinfo{journal}{\bibinfo{title}{Demonstrating multi-round subsystem quantum error correction using matching and maximum likelihood decoders}}.
\newblock {\emph{\JournalTitle{Nature Communications}}} \textbf{\bibinfo{volume}{14}}, \bibinfo{pages}{2852}, \doiprefix\url{10.1038/s41467-023-38247-5} (\bibinfo{year}{2023}).

\bibitem{AcharyaNature2023}
\bibinfo{author}{Acharya, R.} \emph{et~al.}
\newblock \bibinfo{journal}{\bibinfo{title}{Suppressing quantum errors by scaling a surface code logical qubit}}.
\newblock {\emph{\JournalTitle{Nature}}} \textbf{\bibinfo{volume}{614}}, \bibinfo{pages}{676--681}, \doiprefix\url{10.1038/s41586-022-05434-1} (\bibinfo{year}{2023}).

\bibitem{ZhaoPRL2022}
\bibinfo{author}{Zhao, Y.} \emph{et~al.}
\newblock \bibinfo{journal}{\bibinfo{title}{Realization of an error-correcting surface code with superconducting qubits}}.
\newblock {\emph{\JournalTitle{Phys. Rev. Lett.}}} \textbf{\bibinfo{volume}{129}}, \bibinfo{pages}{030501}, \doiprefix\url{10.1103/PhysRevLett.129.030501} (\bibinfo{year}{2022}).

\bibitem{bluvstein2023logical}
\bibinfo{author}{Bluvstein, D.} \emph{et~al.}
\newblock \bibinfo{journal}{\bibinfo{title}{Logical quantum processor based on reconfigurable atom arrays}}.
\newblock {\emph{\JournalTitle{Nature}}} \doiprefix\url{10.1038/s41586-023-06927-3} (\bibinfo{year}{2023}).

\bibitem{gottesman1997stabilizer}
\bibinfo{author}{Gottesman, D.}
\newblock \bibinfo{journal}{\bibinfo{title}{Stabilizer codes and quantum error correction}}.
\newblock {\emph{\JournalTitle{arXiv:quant-ph/9705052}}} \doiprefix\url{10.48550/arXiv.quant-ph/9705052} (\bibinfo{year}{1997}).

\bibitem{ShorCatState}
\bibinfo{author}{Shor, P.~W.}
\newblock \bibinfo{title}{Fault-tolerant quantum computation}.
\newblock FOCS '96, \bibinfo{pages}{56--65}, \doiprefix\url{10.1109/SFCS.1996.548464} (\bibinfo{publisher}{IEEE Computer Society Press}, \bibinfo{year}{1996}).

\bibitem{FowlerSurfaceCodeThresh2009}
\bibinfo{author}{Fowler, A.~G.}, \bibinfo{author}{Stephens, A.~M.} \& \bibinfo{author}{Groszkowski, P.}
\newblock \bibinfo{journal}{\bibinfo{title}{High threshold universal quantum computation on the surface code}}.
\newblock {\emph{\JournalTitle{Phys. Rev. A}}} \textbf{\bibinfo{volume}{80}}, \bibinfo{pages}{052312}, \doiprefix\url{10.1103/PhysRevA.80.052312} (\bibinfo{year}{2009}).

\bibitem{TomitaLowDSC2014}
\bibinfo{author}{Tomita, Y.} \& \bibinfo{author}{Svore, K.~M.}
\newblock \bibinfo{journal}{\bibinfo{title}{Low-distance surface codes under realistic quantum noise}}.
\newblock {\emph{\JournalTitle{Phys. Rev. A}}} \textbf{\bibinfo{volume}{90}}, \bibinfo{pages}{062320}, \doiprefix\url{10.1103/PhysRevA.90.062320} (\bibinfo{year}{2014}).

\bibitem{li2018direct}
\bibinfo{author}{Li, M.}, \bibinfo{author}{Miller, D.} \& \bibinfo{author}{Brown, K.~R.}
\newblock \bibinfo{journal}{\bibinfo{title}{Direct measurement of {B}acon-{S}hor code stabilizers}}.
\newblock {\emph{\JournalTitle{Phys. Rev. A}}} \textbf{\bibinfo{volume}{98}}, \bibinfo{pages}{050301}, \doiprefix\url{10.1103/PhysRevA.98.050301} (\bibinfo{year}{2018}).

\bibitem{ChaoPRXQuantum2020}
\bibinfo{author}{Chao, R.} \& \bibinfo{author}{Reichardt, B.~W.}
\newblock \bibinfo{journal}{\bibinfo{title}{Flag fault-tolerant error correction for any stabilizer code}}.
\newblock {\emph{\JournalTitle{PRX Quantum}}} \textbf{\bibinfo{volume}{1}}, \bibinfo{pages}{010302}, \doiprefix\url{10.1103/PRXQuantum.1.010302} (\bibinfo{year}{2020}).

\bibitem{SteaneSteaneEC1997}
\bibinfo{author}{Steane, A.~M.}
\newblock \bibinfo{journal}{\bibinfo{title}{Active stabilization, quantum computation, and quantum state synthesis}}.
\newblock {\emph{\JournalTitle{Phys. Rev. Lett.}}} \textbf{\bibinfo{volume}{78}}, \bibinfo{pages}{2252--2255}, \doiprefix\url{10.1103/PhysRevLett.78.2252} (\bibinfo{year}{1997}).

\bibitem{KnillKnillEC2005}
\bibinfo{author}{Knill, E.}
\newblock \bibinfo{journal}{\bibinfo{title}{Quantum computing with realistically noisy devices}}.
\newblock {\emph{\JournalTitle{Nature}}} \textbf{\bibinfo{volume}{434}}, \bibinfo{pages}{39--44}, \doiprefix\url{10.1038/nature03350} (\bibinfo{year}{2005}).

\bibitem{HuangPRL2021}
\bibinfo{author}{Huang, S.} \& \bibinfo{author}{Brown, K.~R.}
\newblock \bibinfo{journal}{\bibinfo{title}{Between shor and steane: A unifying construction for measuring error syndromes}}.
\newblock {\emph{\JournalTitle{Phys. Rev. Lett.}}} \textbf{\bibinfo{volume}{127}}, \bibinfo{pages}{090505}, \doiprefix\url{10.1103/PhysRevLett.127.090505} (\bibinfo{year}{2021}).

\bibitem{LaiPRA2017}
\bibinfo{author}{Lai, C.-Y.}, \bibinfo{author}{Zheng, Y.-C.} \& \bibinfo{author}{Brun, T.~A.}
\newblock \bibinfo{journal}{\bibinfo{title}{Fault-tolerant preparation of stabilizer states for quantum calderbank-shor-steane codes by classical error-correcting codes}}.
\newblock {\emph{\JournalTitle{Phys. Rev. A}}} \textbf{\bibinfo{volume}{95}}, \bibinfo{pages}{032339}, \doiprefix\url{10.1103/PhysRevA.95.032339} (\bibinfo{year}{2017}).

\bibitem{BaconBaconShor2006}
\bibinfo{author}{Bacon, D.}
\newblock \bibinfo{journal}{\bibinfo{title}{Operator quantum error-correcting subsystems for self-correcting quantum memories}}.
\newblock {\emph{\JournalTitle{Phys. Rev. A}}} \textbf{\bibinfo{volume}{73}}, \bibinfo{pages}{012340}, \doiprefix\url{10.1103/PhysRevA.73.012340} (\bibinfo{year}{2006}).

\bibitem{aliferis2007subsystem}
\bibinfo{author}{Aliferis, P.} \& \bibinfo{author}{Cross, A.~W.}
\newblock \bibinfo{journal}{\bibinfo{title}{Subsystem fault tolerance with the bacon-shor code}}.
\newblock {\emph{\JournalTitle{Phys. Rev. Lett.}}} \textbf{\bibinfo{volume}{98}}, \bibinfo{pages}{220502}, \doiprefix\url{10.1103/PhysRevLett.98.220502} (\bibinfo{year}{2007}).

\bibitem{PoulinPRL2005}
\bibinfo{author}{Poulin, D.}
\newblock \bibinfo{journal}{\bibinfo{title}{Stabilizer formalism for operator quantum error correction}}.
\newblock {\emph{\JournalTitle{Phys. Rev. Lett.}}} \textbf{\bibinfo{volume}{95}}, \bibinfo{pages}{230504}, \doiprefix\url{10.1103/PhysRevLett.95.230504} (\bibinfo{year}{2005}).

\bibitem{LiBareAnc2017}
\bibinfo{author}{Li, M.}, \bibinfo{author}{Guti\'errez, M.}, \bibinfo{author}{David, S.~E.}, \bibinfo{author}{Hernandez, A.} \& \bibinfo{author}{Brown, K.~R.}
\newblock \bibinfo{journal}{\bibinfo{title}{Fault tolerance with bare ancillary qubits for a [[7,1,3]] code}}.
\newblock {\emph{\JournalTitle{Phys. Rev. A}}} \textbf{\bibinfo{volume}{96}}, \bibinfo{pages}{032341}, \doiprefix\url{10.1103/PhysRevA.96.032341} (\bibinfo{year}{2017}).

\bibitem{LiPRX2019}
\bibinfo{author}{Li, M.}, \bibinfo{author}{Miller, D.}, \bibinfo{author}{Newman, M.}, \bibinfo{author}{Wu, Y.} \& \bibinfo{author}{Brown, K.~R.}
\newblock \bibinfo{journal}{\bibinfo{title}{2d compass codes}}.
\newblock {\emph{\JournalTitle{Phys. Rev. X}}} \textbf{\bibinfo{volume}{9}}, \bibinfo{pages}{021041}, \doiprefix\url{10.1103/PhysRevX.9.021041} (\bibinfo{year}{2019}).

\bibitem{CetinaPRXQuantum2022}
\bibinfo{author}{Cetina, M.} \emph{et~al.}
\newblock \bibinfo{journal}{\bibinfo{title}{Control of transverse motion for quantum gates on individually addressed atomic qubits}}.
\newblock {\emph{\JournalTitle{PRX Quantum}}} \textbf{\bibinfo{volume}{3}}, \bibinfo{pages}{010334}, \doiprefix\url{10.1103/PRXQuantum.3.010334} (\bibinfo{year}{2022}).

\bibitem{DelfosseIEEETransIT2022}
\bibinfo{author}{Delfosse, N.}, \bibinfo{author}{Reichardt, B.~W.} \& \bibinfo{author}{Svore, K.~M.}
\newblock \bibinfo{journal}{\bibinfo{title}{Beyond single-shot fault-tolerant quantum error correction}}.
\newblock {\emph{\JournalTitle{IEEE Transactions on Information Theory}}} \textbf{\bibinfo{volume}{68}}, \bibinfo{pages}{287--301}, \doiprefix\url{10.1109/TIT.2021.3120685} (\bibinfo{year}{2022}).

\bibitem{TansuwannontQuantum2023}
\bibinfo{author}{Tansuwannont, T.}, \bibinfo{author}{Pato, B.} \& \bibinfo{author}{Brown, K.~R.}
\newblock \bibinfo{journal}{\bibinfo{title}{Adaptive syndrome measurements for {S}hor-style error correction}}.
\newblock {\emph{\JournalTitle{Quantum}}} \textbf{\bibinfo{volume}{7}}, \bibinfo{pages}{1075}, \doiprefix\url{10.22331/q-2023-08-08-1075} (\bibinfo{year}{2023}).

\bibitem{chao2018quantum}
\bibinfo{author}{Chao, R.} \& \bibinfo{author}{Reichardt, B.~W.}
\newblock \bibinfo{journal}{\bibinfo{title}{Quantum error correction with only two extra qubits}}.
\newblock {\emph{\JournalTitle{Phys. Rev. Lett.}}} \textbf{\bibinfo{volume}{121}}, \bibinfo{pages}{050502}, \doiprefix\url{10.1103/PhysRevLett.121.050502} (\bibinfo{year}{2018}).

\bibitem{chamberland2018flag}
\bibinfo{author}{Chamberland, C.} \& \bibinfo{author}{Beverland, M.~E.}
\newblock \bibinfo{journal}{\bibinfo{title}{Flag fault-tolerant error correction with arbitrary distance codes}}.
\newblock {\emph{\JournalTitle{Quantum}}} \textbf{\bibinfo{volume}{2}}, \bibinfo{pages}{10--22331} (\bibinfo{year}{2018}).

\bibitem{delfosse2020short}
\bibinfo{author}{Delfosse, N.} \& \bibinfo{author}{Reichardt, B.~W.}
\newblock \bibinfo{journal}{\bibinfo{title}{Short {S}hor-style syndrome sequences}}.
\newblock {\emph{\JournalTitle{arXiv:2008.05051}}} \doiprefix\url{10.48550/arXiv.2008.05051} (\bibinfo{year}{2020}).

\bibitem{LabaziewiczPRL2008}
\bibinfo{author}{Labaziewicz, J.} \emph{et~al.}
\newblock \bibinfo{journal}{\bibinfo{title}{{Suppression of Heating Rates in Cryogenic Surface-Electrode Ion Traps}}}.
\newblock {\emph{\JournalTitle{Physical Review Letters}}} \textbf{\bibinfo{volume}{100}}, \bibinfo{pages}{013001}, \doiprefix\url{10.1103/PhysRevLett.100.013001} (\bibinfo{year}{2008}).

\bibitem{HitePRL2012}
\bibinfo{author}{Hite, D.~A.} \emph{et~al.}
\newblock \bibinfo{journal}{\bibinfo{title}{{100-Fold Reduction of Electric-Field Noise in an Ion Trap Cleaned with In Situ Argon-Ion-Beam Bombardment}}}.
\newblock {\emph{\JournalTitle{Physical Review Letters}}} \textbf{\bibinfo{volume}{109}}, \bibinfo{pages}{103001}, \doiprefix\url{10.1103/PhysRevLett.109.103001} (\bibinfo{year}{2012}).

\bibitem{SedlacekPRA2018}
\bibinfo{author}{Sedlacek, J.~A.} \emph{et~al.}
\newblock \bibinfo{journal}{\bibinfo{title}{{Evidence for multiple mechanisms underlying surface electric-field noise in ion traps}}}.
\newblock {\emph{\JournalTitle{Physical Review A}}} \textbf{\bibinfo{volume}{98}}, \bibinfo{pages}{63430}, \doiprefix\url{10.1103/PhysRevA.98.063430} (\bibinfo{year}{2018}).

\bibitem{Sutherland2022dephasing}
\bibinfo{author}{Sutherland, R.~T.}, \bibinfo{author}{Yu, Q.}, \bibinfo{author}{Beck, K.~M.} \& \bibinfo{author}{H\"affner, H.}
\newblock \bibinfo{journal}{\bibinfo{title}{One- and two-qubit gate infidelities due to motional errors in trapped ions and electrons}}.
\newblock {\emph{\JournalTitle{Phys. Rev. A}}} \textbf{\bibinfo{volume}{105}}, \bibinfo{pages}{022437}, \doiprefix\url{10.1103/PhysRevA.105.022437} (\bibinfo{year}{2022}).

\bibitem{AliferisQIC2007}
\bibinfo{author}{Aliferis, P.} \& \bibinfo{author}{Terhal, B.~M.}
\newblock \bibinfo{journal}{\bibinfo{title}{Fault-tolerant quantum computation for local leakage faults}}.
\newblock {\emph{\JournalTitle{Quantum Info. Comput.}}} \textbf{\bibinfo{volume}{7}}, \bibinfo{pages}{139–156}, \doiprefix\url{10.26421/QIC7.1-2-9} (\bibinfo{year}{2007}).

\bibitem{KielpinskiQCCD2002}
\bibinfo{author}{Kielpinski, D.}, \bibinfo{author}{Monroe, C.} \& \bibinfo{author}{Wineland, D.~J.}
\newblock \bibinfo{journal}{\bibinfo{title}{Architecture for a large-scale ion-trap quantum computer}}.
\newblock {\emph{\JournalTitle{Nature}}} \textbf{\bibinfo{volume}{417}}, \bibinfo{pages}{709--711}, \doiprefix\url{10.1038/nature00784} (\bibinfo{year}{2002}).

\bibitem{TaylorNatPhys2005}
\bibinfo{author}{Taylor, J.} \emph{et~al.}
\newblock \bibinfo{journal}{\bibinfo{title}{Fault-tolerant architecture for quantum computation using electrically controlled semiconductor spins}}.
\newblock {\emph{\JournalTitle{Nature Physics}}} \textbf{\bibinfo{volume}{1}}, \bibinfo{pages}{177--183}, \doiprefix\url{10.1038/nphys174} (\bibinfo{year}{2005}).

\bibitem{KangPRAp2023}
\bibinfo{author}{Kang, M.} \emph{et~al.}
\newblock \bibinfo{journal}{\bibinfo{title}{Designing filter functions of frequency-modulated pulses for high-fidelity two-qubit gates in ion chains}}.
\newblock {\emph{\JournalTitle{Phys. Rev. Appl.}}} \textbf{\bibinfo{volume}{19}}, \bibinfo{pages}{014014}, \doiprefix\url{10.1103/PhysRevApplied.19.014014} (\bibinfo{year}{2023}).

\bibitem{brown2004arbitrarily}
\bibinfo{author}{Brown, K.~R.}, \bibinfo{author}{Harrow, A.~W.} \& \bibinfo{author}{Chuang, I.~L.}
\newblock \bibinfo{journal}{\bibinfo{title}{Arbitrarily accurate composite pulse sequences}}.
\newblock {\emph{\JournalTitle{Physical Review A}}} \textbf{\bibinfo{volume}{70}}, \bibinfo{pages}{052318}, \doiprefix\url{10.1103/PhysRevA.70.052318} (\bibinfo{year}{2004}).

\bibitem{Wang2020hifi}
\bibinfo{author}{Wang, Y.} \emph{et~al.}
\newblock \bibinfo{journal}{\bibinfo{title}{High-fidelity two-qubit gates using a microelectromechanical-system-based beam steering system for individual qubit addressing}}.
\newblock {\emph{\JournalTitle{Phys. Rev. Lett.}}} \textbf{\bibinfo{volume}{125}}, \bibinfo{pages}{150505}, \doiprefix\url{10.1103/PhysRevLett.125.150505} (\bibinfo{year}{2020}).

\end{thebibliography}

\end{document}